\documentclass[10pt,conference]{IEEEtran}
\usepackage[utf8]{inputenc}
\usepackage{cite}
\usepackage{amsmath,amssymb,amsfonts}
\usepackage{algorithmic}
\usepackage{graphicx}
\usepackage{textcomp}
\usepackage{xcolor}

\usepackage{dsfont}
\usepackage{hyperref}
\usepackage{booktabs}
\usepackage{multirow}
\usepackage{subcaption}
\usepackage[numbers]{natbib}
\usepackage{enumitem}

\begin{document}

\title{PYInfer: Deep Learning Semantic Type Inference for Python Variables
}

\author{\IEEEauthorblockN{Siwei Cui}
\IEEEauthorblockA{
\textit{Texas A\&M University}\\
College Station, Texas \\
siweicui@tamu.edu}
\\
\IEEEauthorblockN{Luochao Wang}
\IEEEauthorblockA{
\textit{Texas A\&M University}\\
College Station, Texas \\
wangluochao@tamu.edu}
\and
\IEEEauthorblockN{Gang Zhao}
\IEEEauthorblockA{
\textit{Texas A\&M University}\\
College Station, Texas \\
zhaogang92@tamu.edu}
\\

\IEEEauthorblockN{Ruihong Huang}
\IEEEauthorblockA{
\textit{Texas A\&M University}\\
College Station, Texas \\
huangrh@cse.tamu.edu}
\and
\IEEEauthorblockN{Zeyu Dai}
\IEEEauthorblockA{
\textit{Texas A\&M University}\\
College Station, Texas \\
jzdaizeyu@tamu.edu}
\\
\IEEEauthorblockN{Jeff Huang}
\IEEEauthorblockA{
\textit{Texas A\&M University}\\
College Station, Texas \\
jeffhuang@tamu.edu}

}

\maketitle

\begin{abstract}
Python type inference is challenging in practice.
Due to its dynamic properties and extensive dependencies on third-party libraries without type annotations,
the performance of traditional static analysis techniques is limited.
Although semantics in source code can help manifest intended usage for variables (thus help infer types),
they are usually ignored by existing tools.
In this paper, we propose PYInfer, an end-to-end learning-based type inference tool that
automatically generates type annotations for Python variables. 
The key insight is that contextual code semantics is critical in inferring the
type for a variable. For each use of a variable, we collect a few tokens
within its contextual scope, and design a neural network to predict its
type. One challenge is that it is difficult to collect a high-quality
human-labeled training dataset for this purpose. To address this issue, we apply
an existing static analyzer to generate the ground truth for
variables in source code.

Our main contribution is a novel approach to statically infer variable types effectively and efficiently. 
Formulating the type inference as a classification
problem, we can handle user-defined types and predict type probabilities for
each variable. Our model achieves 91.2\% accuracy on classifying 11 basic types in
Python and 81.2\% accuracy on classifying 500 most
common types. Our results substantially
outperform the state-of-the-art type annotators. Moreover, PYInfer achieves
5.2X more code coverage and is 187X faster than a state-of-the-art learning-based
tool. With similar time consumption, our model annotates 5X more variables
than a state-of-the-art static analysis tool (PySonar2). 
Our model also outperforms a learning-based function-level annotator (TypeWriter)
on annotating types for variables and function arguments.
All our tools and datasets are publicly available to facilitate future research in this direction.
\end{abstract}


\section{Introduction}

Python is widely used due to its flexibility and the abundance of third-party
libraries (e.g., web and machine learning frameworks). However, flexibility
brings challenges to code optimization and also makes it error-prone. Variable type
inconsistency is a common error in dynamic languages. Due to Python's dynamic property, 
the interpreter cannot check type inconsistency as a
static programming language compiler (e.g., Go or Rust). 
Python type checkers~\cite{mypy,pyre-check,pytype,pyright} 
take advantage of annotations to detect type inconsistencies. 
These tools primarily rely on manually written type annotations by developers,
which are expensive to provide. 

To facilitate user programming and checking type errors, variable type
inference is a necessary step. 
Deep learning has been applied to infer types for 
JavaScript~\cite{hellendoorn2018deep,malik2019nl2type, pradel2018deepbugs,wei2020lambdanet} 
by leveraging TypeScript~\cite{bierman2014understanding} to 
generate large corpus with precise annotations. 
However, there exist few good solutions for Python because of its broad scope of
dynamic features and extensive dependencies on third-party libraries,
which left us many opportunities in this scope. 
The quality of the dataset itself brings a large gap between annotating Python and JavaScript.

Type inference tools that apply static
analysis or dynamic analysis~\cite{cannon2005localized, salib2004starkiller,
vitousek2014design} do not require labeled annotations for type inference. 
However, they are imprecise and leave out the abundant natural language semantics in source code.
\citet{xu2016python} proposed to use a probabilistic model to infer variable types by
leveraging type hints from data flow, attributes, subtypes, and variable names.
However, it takes considerable time to analyze the source code and solve the probabilistic constraints.
Without a sufficiently large dataset to provide enough signals, 
the performance of the probabilistic model is also limited.

Existing human-labeled type annotations in mypy~\cite{mypy} and typeshed~\cite{typeshed} 
only cover few annotations for function arguments and return types.
The dataset contains no variable annotation, and is inadequate for inferring Python variable types. 
TypeWriter~\cite{pradel2019typewriter} also targets type inference for Python, 
but it addresses the problem of inferring function arguments and return types.
Experiments show that TypeWriter is insufficient on variable type prediction.
The function level annotations (function arguments and return types)
 are useful to serve as an API contract for IDEs, 
while variable level annotations can be used to provide type check for each variable.


In this paper, we present PYInfer, a deep learning based approach to generate
type annotations for Python.
A high-level overview of PYInfer is depicted in Fig.~\ref{fig:SmallModel}. 
Since the human-labeled dataset for variables is not available, 
we first employ a static analyzer,
PySonar2~\cite{pysonar2}, to automatically generate initial annotations from top-star Python GitHub projects.
We then apply a series of data cleaning techniques to refine the quality of our dataset.
We further feed the annotations and contextual information to
train a deep neural network, which ranks each type with probabilities effectively. 

We highlight that fusing deep learning with static
analysis to infer type annotations is promising. By combining deep
learning with static analysis from end to end, our approach is capable of
analyzing code semantics with well-developed natural language
processing (NLP) techniques. 
PYInfer's effectiveness benefits from addressing the following challenges:

\begin{figure}[!htb]
  \includegraphics[width=\linewidth]{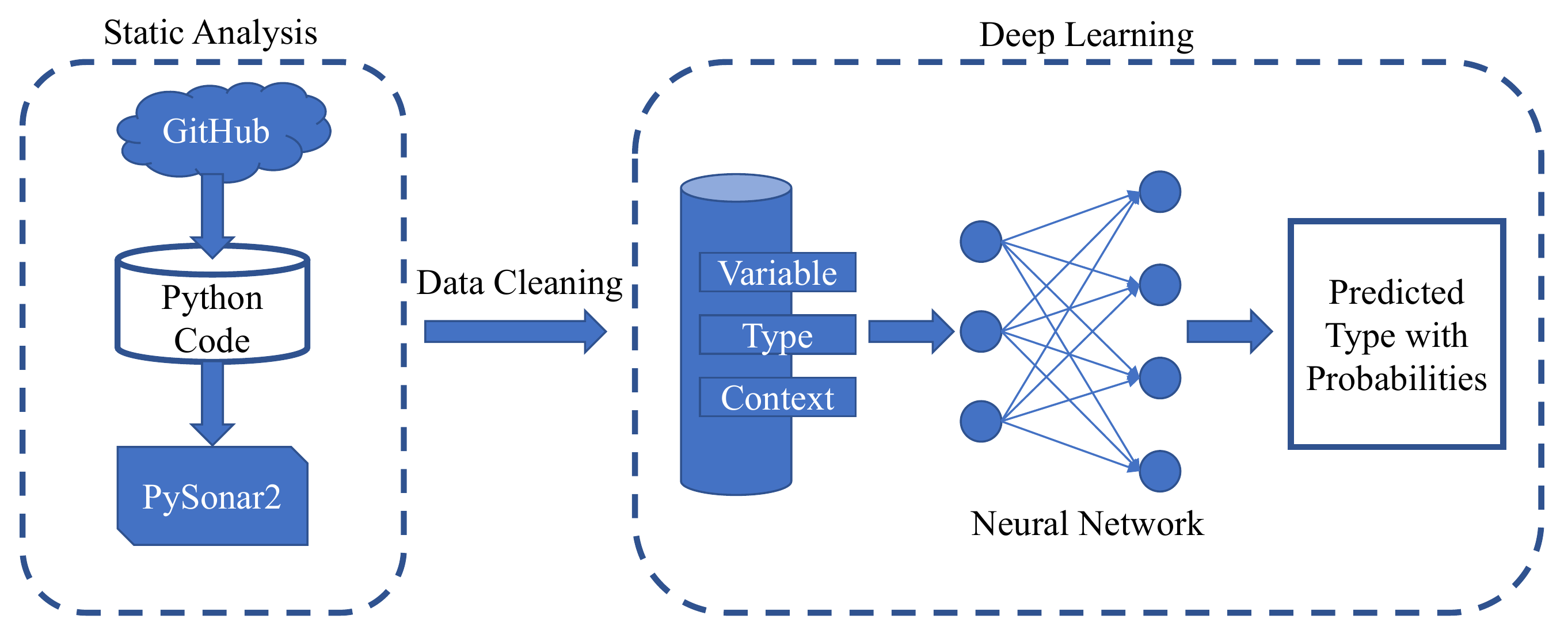}
  \caption{A high-level overview of PYInfer. We collect Python source
  code from 4,577 top-star GitHub repositories, apply
  PySonar2 to generate initial variables annotations, and encode a
  contextual neural network for variable type prediction.}
  \label{fig:SmallModel}
\end{figure}

\begin{figure}[!htb]
  \includegraphics[width=\linewidth]{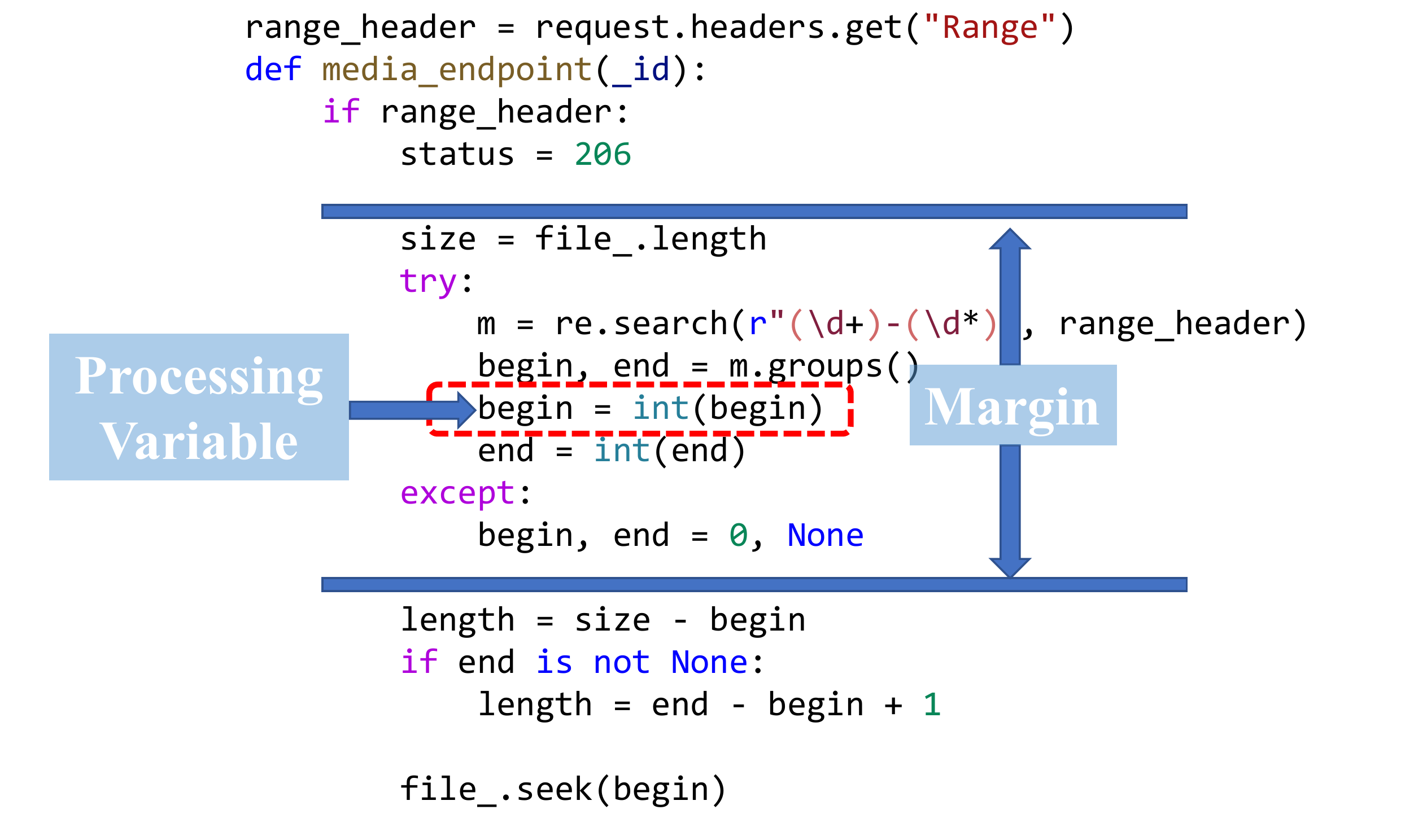}
  \caption{An illustration of using margin to characterize contextual semantics for variable type inference.}
  \label{fig:margin}
\end{figure}

\begin{itemize}[leftmargin=*]

  \item \textbf{Annotation Dataset Collection.}
  Analyzing contextual source code semantics for variable types demands a large annotated dataset.
  However, there exists no well-acknowledged large dataset with annotations.
  We generate our annotated dataset with enriched data based on inference results from PySonar2 
and perform data cleaning to enhance quality,
 which itself is a significant contribution, because high quality labeled data is critical for deep learning. 
  Our dataset is collected from 4,577 popular Python projects on GitHub with 54,928,617 Lines
  Of Code (LoC) from 320,402 source files. It contains 77,089,946 annotations,
   which is large enough for most Python types research. 

  \item \textbf{User-defined Types.} Due to the flexibility of Python, types can
  be user-defined and changed during runtime. We frame Python type
  inference as a classification task to cover user-defined types in our 500 most common types.
  We investigate the performance of 11 basic types compared with 500 types.
  Our model achieves 91.187\% accuracy on classifying 11 basic types and 81.195\% on predicting 500 types.
  This is a significant advance over past work~\cite{xu2016python} on this problem.
  As a classification task, our model provides confidence levels for each type.
  It achieves 97.677\% precise with the 0.9 threshold on the confidence level.

  \item \textbf{Source Code Embeddings.} Source code contains abundant semantic
  information and type hints in variable names and usages, which is helpful for type inference.
  Previous work~\cite{xu2016python,malik2019nl2type,pradel2019typewriter} 
  has applied word embeddings for type inference. 
  However, we show that these embeddings do not work well due to the Out-Of-Vocabulary (OOV)
  issue caused by the large number of dynamic features and user-defined types in Python.
  To tackle this problem, we employ the Byte Pair Encoding (BPE)~\cite{sennrich2015neural} algorithm. 
  It provides sufficient signals to analyze semantics in variable names and contextual data.
  Compared with graph-based embeddings in LambdaNet~\cite{wei2020lambdanet}, 
  BPE embeddings are lightweight and can be easily extended to analyze other languages.
  We demonstrate that BPE is effective in inferring variable types.
  Our model improves 27.1\% accuracy with BPE embeddings over GloVe embeddings~\cite{pennington2014glove}.

  \item \textbf{Contextual Code Semantics.} A key insight in our approach is
  leveraging contextual code semantics for variable type inference. 
  We hypothesize that the context within a certain margin conveys relevant semantic
   information to characterize the variable.
  Inspired by interprocedural static analysis~\cite{cannon2005localized, salib2004starkiller, vitousek2014design},
  our approach is capable of analyzing the semantics of variables 
   together with the structural syntax and grammar information. 
   The setting of the margin hyperparameter is illustrated in Fig.~\ref{fig:margin}.
   For each variable, we collect source code tokens within its contextual scope.
   We adopt the Gated Recurrent Unit (GRU)~\cite{cho2014learning} with the
    attention mechanism~\cite{vaswani2017attention} to analyze contextual semantics.
     Our ablation test on contextual information show 41.0\% improvement on accuracy. 
  Our evaluation on human-labeled typeshed~\cite{typeshed} dataset demonstrates the same result.
  The contextual information provides local semantics for variables,
   and it is useful for deriving variable annotations. 

\end{itemize}

Putting all these contributions together, we develop an end-to-end,
 highly effective and efficient framework to infer variable types for Python statically.
Our dataset is large enough for most research under Python types, which itself is a novel contribution.
We achieve the accuracy of 91.187\% on 11 basic types, and 81.195\% on 500 most common types. 
PYInfer demonstrates superiority in both coverage and time efficiency on large 
projects compared to existing work~\cite{xu2016python}. 
Instead of assigning weights to multiple factors for probabilistic inference~\cite{xu2016python},
PYInfer achieves 5.2X more coverage and is 187.4X faster. 
Our model annotates a variable on an average of 1 millisecond.
Compared with PySonar2, 
our tool takes a similar time on analysis but generates 5X more annotations. 
A motivating example comparing PYInfer and PySonar2 is provided in Section~\ref{Discussion}.
Although trained on the annotations generated by PySonar2,
PYInfer can handle sophisticated cases using contextual code semantics,
as shown in Fig.~\ref{fig:PySonar2VSPYInfer}.
PYInfer can also be extended to perform function argument inference.
It shows superiority over TypeWriter on inferring variable types and function arguments.

\begin{figure*}[!htb]
  \includegraphics[width=\linewidth]{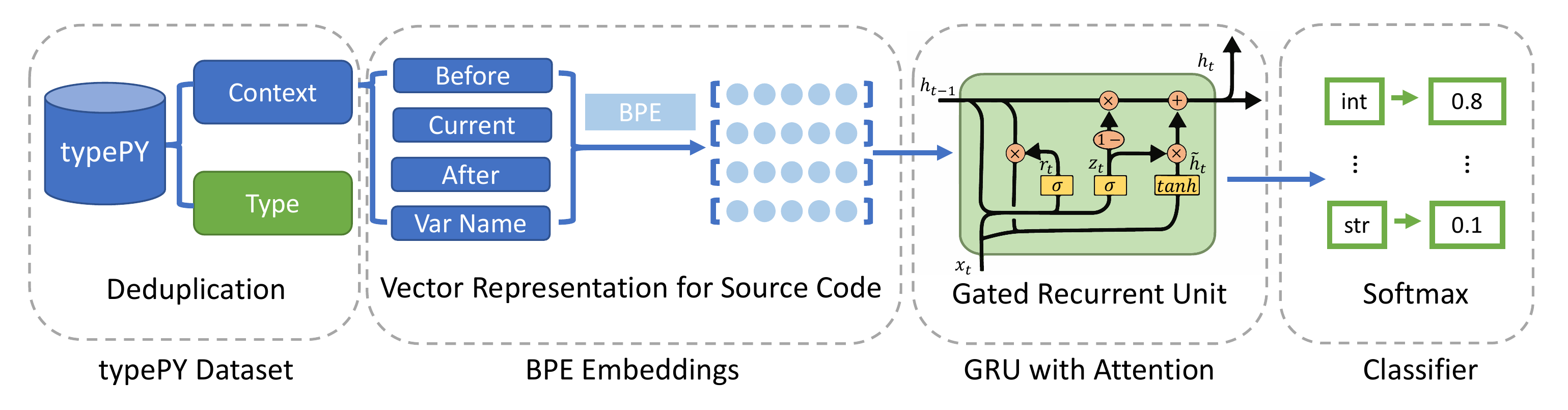}
  \caption{PYInfer Model Framework.}
  \label{fig:Model}
\end{figure*}

We have released our PYInfer model, source code, and dataset to facilitate
further research\footnote{The link will be provided after the
double-blind review process.}.
Existing type checkers can benefit from the type annotations generated by PYInfer to detect type inconsistencies. 
We provide a workflow on integrating PYInfer with pyre to detect variable inconsistencies in Python repositories.
As an end-to-end static type annotator, PYInfer provides annotations with probabilities in 1 ms 
for a variable. 
It provides the type annotation for programmers seamlessly while they are programming.
Our framework can also be used to infer argument types.
Since our approach relies on high-level semantics rather than graph structures, 
it can be easily extended into annotating variables and detecting semantic errors in
other dynamically typed languages.

\section{PYInfer Framework}

Fig.~\ref{fig:Model} presents a
technical overview of the PYInfer model. The basic idea
is to formulate the type inference as a classification problem. We examine the top 500
types based on the frequency of occurrence, and analyze the contextual
semantics within a certain margin. 
The BPE algorithm is adopted to derive the vector representation as
embeddings. We further feed these embeddings into a GRU network with the attention mechanism to extract
code semantics from the context. We then employ a softmax layer to classify each type with
probabilities. PYInfer consists of the following four components: 
data collection and generation, source code embeddings, model formulation, and model training.

\subsection{Data Collection and Generation}  \label{dataset}

To classify variable types,
we need a sufficiently large annotated dataset. 
As the human-labeled variable annotation dataset is unavailable, we adopt
PySonar2 to generate the initial annotations. 
Since PySonar2 does conservative analysis, we ignore all variables for which it cannot make type inference,
and assume the results for the rest variables are ground truth (though the types may be over-approximated). 
We also analyze typeshed~\cite{typeshed}, probPY~\cite{xu2016python},
 and TypeWriter~\cite{pradel2019typewriter} dataset.

Our datasets are summarized in Table~\ref{tab:dataset}. The \emph{Original} and \emph{Valid} present 
the number of annotations before and after data cleaning and deduplication.
\emph{typePY} is our source code dataset collected from 4,577 top-star GitHub repositories.
\emph{probPY} stands for the dataset published in~\citet{xu2016python}, and
\emph{typeshed} is a human-labeled dataset containing only the annotations
 for function parameters and return values~\cite{typeshed}.

We collect the \emph{typePY} dataset by annotating Python source code in top-star GitHub repositories.
For each variable, we store the link to the repository, 
file name, variable name, start and end token location, type annotation, and corresponding source code. 
To obtain type annotations, we adopt the PySonar2, an advanced semantic indexer
for Python~\cite{pysonar2}, to infer types for each variable. 
We obtain 77,089,946 annotations from 320,402 Python source code files.
As PySonar2 has only 49.47\% coverage on annotations~\cite{xu2016python},
we also refine our dataset with a series of data cleaning techniques.
We eliminate all meaningless types, such as ``question marks'' and ``None'' types,
and perform deduplication.   
Eventually, we obtain 42,560,876 valid annotations. 
To the best of our knowledge, 
this is the largest Python annotation dataset. We release this dataset to facilitate research on this topic.

\begin{table}[!htb]
  \caption{Metadata for typePY, probPY and typeshed dataset. }
  \label{tab:dataset}
  \setlength{\tabcolsep}{0.8mm}{
  \begin{tabular}{cccccc}
  \toprule
  \multicolumn{2}{c}{\textbf{Dataset}}                       & \textbf{Python Files} & \textbf{LoC} & \textbf{Original} & \textbf{Valid} \\ \midrule
  \multicolumn{2}{c}{\textbf{typePY}}                        & 320,402               & 54,928,617   & 77,089,946        & 42,560,876     \\ \midrule
  \multicolumn{2}{c}{\textbf{probPY dataset}}                & 716                   & 146,019      & 106,808           & 64,831         \\ \midrule
  \multicolumn{1}{c|}{\multirow{2}{*}{\textbf{typeshed}}} & \textbf{stdlib}       & 545                   & 34,854       & 38,536            & 14,685         \\
  \multicolumn{1}{c|}{}        & \textbf{third-party} & 564                   & 23,877       & 17,632            & 8,957          \\ \bottomrule
  \end{tabular}}
  \end{table}

For the \emph{probPY}~\cite{xu2016python} dataset, we exploit the merged data
combining the result of PySonar2 and dynamic analysis. 
The \emph{probPY} dataset provides the variable name,
annotation, and the source code to generate contextual information for our
model. We are able to compare the
time consumption and coverage between PYInfer, PySonar2, and probPY
model.

The \emph{typeshed} dataset contains human-label type annotations for Python
standard libraries and third-party packages, making it more reliable.
However, the typeshed dataset only covers the annotations for function parameters and return types, and
there is no contextual information due to frequent code updates. 
We extract the argument annotations to evaluate our model for the importance of contextual code semantics. 
After merging the annotations in third-party libraries with standard libraries, 
we deduplicate annotations on the [variable name, type annotation] pairs. 
We finally obtain 3,379 different types with 16,537 annotations in this dataset. 
Those more than 30.1\% duplicates indicate that programmers are likely to apply similar 
names on variables with similar functionalities.

\subsection{Source Code Embeddings}

The conventional embeddings approach builds up a dictionary on frequently used tokens, 
and generates embeddings for every token in the vocabulary. 
This approach guarantees that every token keeps intact when feeding to the model. 
Many existing approaches, such as Word2Vec~\cite{le2014distributed} and Global Vectors for 
word representation (GloVe)~\cite{pennington2014glove}, 
retrieve the vectorization representation of each token. 
However, word embedding approaches are not suitable for source code. 
With plenty of user-defined variable names and function names, 
these approaches often suffer from OOV as rare words are not observed in training data. 
The OOV makes the model unable to absorb any information from tokens not in the vocabulary. 
A trivial solution is to use $<$UNK$>$ to represent unknown words, 
which is inappropriate due to the loss of semantic meanings.

To capture those semantics in variable and function names, we adopt the 
BPE~\cite{sennrich2015neural} algorithm to generate embeddings for source code.  
The BPE algorithm is first known as a compression
algorithm~\cite{gage1994new,shibata1999byte} and is effective in 
many program analysis tasks with neural
networks~\cite{sennrich2015neural,babii2019modeling,devlin2018bert,karampatsis2019maybe,karampatsis2020big}. 
This algorithm alleviates the OOV by merging a most frequent byte pair into a new byte.
Start with a single individual character, we characterize user-defined tokens by 
splitting them into smaller pieces in our dictionary.
We adopt the bottom-up clustering approach, and initially generate unigram for all characters. 
After that, we iteratively calculate the frequencies of n-gram and 
utilize a greedy approximation to maximizing the language model log-likelihood to 
generate a new n-gram for most frequent n-gram pairs~\cite{sentencepiece}. 

We train our BPE model on source code corpus and obtain 19,995 different base words for embedding generation. 
Compared with conventional embedding approaches, 
BPE embeddings make full use of contextual code semantics by resolve OOV. 
We demonstrate that BPE is effective in embedding Python source code for variable annotations. 
Variable names using snake case (e.g., network\_address) or camel case can be effectively embedded using BPE.

\subsection{Model Formulation}

One of the key insights is that we bring the contextual information to our model. 
The contexts in source code not only carry meaningful semantic knowledge but also delivers 
clues and insights of variable functionalities. We set up the margin $m$ to indicate how much
contextual information will be considered.
The setting of the margin is illustrated in Fig.~\ref{fig:margin}.
For each variable $v_i$, $i \in [1..n]$, we process the current line possessing the current variable, 
which is annotated as $d_i^c$, $m$ tokens before the current line, 
as annotated with $d_i^e$, $m$ tokens after the current line, which is represented as $d_i^a$, 
and the name of the current variable $d_i^n$. The contextual information provides local semantics for variables,
which is adequate to derive variable annotations. 
Let the BPE algorithm be $\mathcal{B}()$, 
and we can finalize our embedding $\mathcal{X}_i$ for variable $v_i$ as:
$$\mathcal{X}_i= \mathcal{B}(d_i^e)+\mathcal{B}(d_i^c)+\mathcal{B}(d_i^a)+\mathcal{B}(d_i^n),$$

where the ``+'' standards for the concatenation on two embedding vectors. 
We deliberately assign the embeddings of the variable name in the latter part in the embeddings. 
This setting enables us to obtain the semantic representation by extracting 
the pattern in the final layer in our GRU model.

To characterize embedding features comprehensively, we adopt the Gated Recurrent Unit 
(GRU)~\cite{cho2014learning}, a recurrent neural network (RNN)~\cite{schuster1997bidirectional} 
which has similar performance compared to bidirectional recurrent neural 
networks~\cite{schuster1997bidirectional} but with lower computational complexity. 
GRU is used to analyze and characterize the embedding of our source code in a superior manner. 
For each variable, GRU is adopted to analyze the characterization of contextual information 
combined with the variable names. It is capable of processing a sequence of a vector by applying 
a recurrence formula at every time step $t$. Initially, we have the output vector $h_0=0$ when $t=0$. 
Suppose the number of the tokens after embedding is $s_i$ for variable $v_i$. 
For each input source code token $x_t$, where $t \in [1..s_i]$ in embeddings $\mathcal{X}_i$, we have:
\[
  \begin{split}
    z_t &= \sigma(W_z \cdot x_t + U_z \cdot h_{t-1} + b_z), \\
    r_t &= \sigma(W_r\cdot x_t + U_r \cdot h_{t-1} + b_z), \\
    h_t &= z_t \odot h_{t-1} + (1-z_t) \odot \phi( W_h\cdot x_t + U_h(r_t \odot h_{t-1}) + b_h),
  \end{split}
\]

where $x_t$ is our input embeddings with contextual information, 
$h_t$ is our output vector for this variable $v_i$, $z_t$ this the update gate and $r_t$ is the reset gate, 
and $W, U$ and $b$ are parameters in our model. The $\sigma$ stands for the sigmoid function, 
and $\phi$ represents the hyperbolic tangent. 
The attention mechanism~\cite{vaswani2017attention} is also added to our model for more reliable performance. 
Since variable names are appended in our embedding vector's last position, 
we can extract the final layer in our output vector $h_{s_i}$ to characterize variable $\mathcal{X}_i$. 
A dropout layer is attached to the GRU model to address the overfitting problem properly. 
Finally, We attach a fully connected layer to the output of the GRU model to give our model more flexibility 
on learning contextual code semantics.


We perceive the Python type inference as a type classification to address user-defined types. 
To obtain the probability of each type, we apply the softmax regression (also known as multinomial 
logistic regression~\cite{bohning1992multinomial}) on the feature extracted from the output of the GRU model. 
For the output of GRU network $h_{s_i}$, we have:
$$y_i = argmax(P(h_{s_i})) = argmax(\frac{e^{h_{s_i}}}{\sum_{i=0}^n{e^{h_{s_i}}}}),$$

where $argmax$ is a function returns the position with the maximum probability, 
and $P(h_{s_i})$ is a list of probabilities for each possible type of variable $v_i$. 
This function is used to approximate a target integer $y_i\in [1,\mathcal{C}]$, 
where $\mathcal{C}$ represents the number of classes. 
The softmax function $P()$ produces a scalar output $P(h_{s_i})\in \mathds{R}$, 
with the probability for each type $P(h_{s_i})_j \in [0,1]$.

With the help of the softmax layer, we are able to generate the type annotations for each variable
with the distribution on probabilities $P(h_{s_i})$. 
PYInfer annotates variable based on the type with the maximum probability, which returns $y_i$. 
We are able to add a threshold on the confidence level, i.e., the probability. Our model accuracy
increases with the rise of the confidence level (Table~\ref{tab:threshold}), 
which demonstrates the effectiveness of our model on types classification.

\subsection{Model Training}

In our model, we adopt the cross-entropy as our loss function. 
Specifically, we apply a softmax followed by a logarithm to derive the confidence for the type inference,
 and append the negative log-likelihood loss (NLLLoss) on the softmax results. 
 For the loss function, we have,

$$\mathcal{L} = \sum_{i=1}^n{\sum_{j=1}^\mathcal{C}{-g_{ij}*log(P(h_{s_i})_j)}},$$

where $\mathcal{L}$ stands for the loss function for our model, 
which is calculated by summing up the loss for each annotation. 
The constant $n$ is the total number of annotations we have, 
and $\mathcal{C}$ represents the number of classes. 
The ground truth type for variable $v_i$ is represented by $g_{ij}$. We set $g_{ij} = 1$ if 
the annotation for variable $v_i$ is $j$, and $g_{ij} = 0$ otherwise. 
$P(h_{s_i})_j$ defines the log softmax result of class $j$ on variable $v_i$.

We further feed the computational results into an optimizer, 
which aims at minimizing the value of the loss function $\mathcal{L}$. 
Extensive experiments are conducted, 
and Adam Optimizer~\cite{kingma2014adam} turns out to be the best fit for this phenomenon. 

\section{Evaluation} \label{Evaluation}

In this section, we evaluate PYInfer by answering the
following research questions:
\begin{itemize}
  \item RQ 1: How effective is PYInfer at deriving the correct type annotations?
  \item RQ 2: Does the number of classes considered for classification have a significant impact on PYInfer?
  \item RQ 3: How does the threshold influence our PYInfer model?
\end{itemize}



\subsection{RQ 1: Model Effectiveness and Baseline Comparison}

\subsubsection{Dataset and Experiment Settings}

In this RQ, we analyze our context model's performance with 500 most common types in the typePY dataset. 
Besides all built-in types in Python, 
we also consider a large amount of user-defined types. 
After analyzing the source code corpus, we find that duplicates exist in our dataset. 
The main reason is that some GitHub repositories reuse the same code from the others. 
We perform deduplication on our dataset, and derive 3,499,933 annotations for evaluation.
We also inspect some of the type distribution, as shown in Table~\ref{tab:P1DataDist}. 
This uneven distribution reflects the difference in real-world variable usage on different types. 
Our whole data corpus is randomly split into training, validation, 
and testing data with the proportion of 60\%:20\%:20\%. We run all experiments on a single machine with Intel i7-9700k CPU,
32GB RAM, and a single NVIDIA RTX 2070 Super GPU.

\begin{table}[!htb]
  \caption{RQ 1: Type annotation distribution of 500 most common types in typePY dataset.}
  \label{tab:P1DataDist}
  \begin{tabular}{cc|cc}
    \toprule
  \textbf{Type}         & \textbf{Annotations Count} & \textbf{Type}        & \textbf{Annotations Count} \\ \midrule
  str                   & 921,471                     & DataFrame            & 16,764                      \\
  int                   & 628,552                     & Series               & 13,429                      \\
  dict                  & 336,374                     & {[}float{]}          & 11,376                      \\
  bool                  & 241,804                     & tuple                & 11,110                      \\
  float                 & 126,005                     & (int, int)           & 9,738                       \\
  {[}str{]}             & 121,669                     & \{dict  dict\}       & 8,608                       \\
  list                  & 48,400                      & object               & 6,110                       \\ \bottomrule
  \end{tabular}
  \end{table}

\subsubsection{Implementing Details and Results}

We train our PYInfer model with parameters in Table~\ref{tab:P1Param}, 
and report testing results. To analyze contextual code semantics, 
we generate the embeddings by analyzing contextual semantics in source code with a separator 
between each part in our margin settings. 
We obtain the vector representations by extracting the final layer in the GRU neural network
as the variable names are in the latter part in embeddings.

We add a dropout layer to tackle the overfitting problem, making our model more 
generalizable and elevate performance for real-world cases. 
The parameters MODEL\_SIZE and SEQ\_LEN characterize the size of the hidden layer in the GRU network. 
We set up hyperparameter TENSOR\_LEN to eliminate some extreme long embeddings, 
which is often the case where a piece of source code contains immense unseen tokens. 
We have collected the distribution on the embedding length, 
and the embedding length within 1,000 covers 99.9\% annotations in our dataset. 
Therefore, we can safely adopt the annotations within this length constrain to train our model.

\begin{table}[!htb]
  \centering
  \caption{Hyperparameters for PYInfer model, number of annotations, and testing results.}
  \label{tab:P1Param}
  \setlength{\tabcolsep}{0.6mm}{
  \begin{tabular}{cccc}
  \toprule
  \textbf{Optimizer}   & \textbf{Margin}         & \textbf{Loss}   & \textbf{Dropout}   \\ \hline
  Adam & 128    & CrossEntropy  &  0.1  \\ \hline
  \textbf{Learning\_Rate} & \textbf{MODEL\_SIZE} & \textbf{SEQ\_LEN} & \textbf{TENSOR\_LEN} \\ \hline
  512     & 1,000 & 0.0001  & 512\\ \toprule
  \textbf{Training Samples} & \textbf{Validation Samples} & \textbf{Testing Samples}  \\ \hline
   2,099,739 & 699,913 & 699,914  \\  \toprule
   \multicolumn{4}{c}{\textbf{PYInfer Model Testing Results}}  \\ \hline
  \textbf{Accuracy} & \textbf{Precision} & \textbf{Recall} & \textbf{F-1 Score} \\ \hline
  81.195\% & 79.318\% & 81.195\% & 80.246\% \\  \bottomrule

  \end{tabular}}
  \end{table}

With all the settings above, we fine-tune the parameters and employ the accuracy as
 one of our model's evaluation matrices. The accuracy is calculated as:

$$A_{\mathcal{X}_i}=\frac{|\mathcal{D}({\mathcal{X}_i})\wedge \mathcal{C}({\mathcal{X}_i})|}{|\mathcal{D}({\mathcal{X}_i})|},$$

where ${\mathcal{X}_i}$ represents the current embeddings being processed,
 $\mathcal{D}({\mathcal{X}_i})$ represents the ground-truth type for variable $v_i$, 
 and $\mathcal{C}({\mathcal{X}_i})$ returns the Top-1 type annotation ranked based on the
  probability. Our model eventually achieves 81.195\% accuracy on testing data.

Since the distribution for each type is uneven, we also evaluate our model 
using weighted precision and recall, and calculate the f-1 score based on them. 
As a multi-class classification, we calculate the average of each evaluation matrix weighted by support,
 i.e., the number of correct annotations for each type. 
 PYInfer achieves 79.318\% on precision, 81.195\% on recall, and 80.246\% on the f-1 score.

\subsubsection{Baseline Analysis and Insights}
\label{sec:baseline}

For baseline analysis, we compare our PYInfer framework with the probPY model~\cite{xu2016python} and PySonar2.
First, we perform baseline experiments using the merged data in the probPY dataset. 

Evaluation matrices in probPY take advantage of the recall and precision. 
The precision in probPY is calculated based on the feasible input. However, 
due to the lack of ground truth, they provide estimations based on randomly selected samples. 
Therefore, we adopt the recall for accurate evaluation. 
Similar to probPY, we calculate the recall as:
$$R_{\mathcal{X}_i}=\frac{|\mathcal{D}({\mathcal{X}_i})\wedge \mathcal{C}({\mathcal{X}_i},TOP_k)|}{|\mathcal{D}({\mathcal{X}_i})|}$$

The ${\mathcal{X}_i}$ is the embeddings of variable $v_i$ being processed.
 $\mathcal{D}({\mathcal{X}_i})$ represents the ground-truth type for 
variable $v_i$, and $\mathcal{C}({\mathcal{X}_i},TOP_k)$ returns the $TOP_k$ type annotations 
ranked based on the probabilities for each possible type.

\begin{table}[!htb]
  \centering
  \caption{Evaluation results on recall using probPY dataset.}
  \label{tab:xu16}
  \begin{tabular}{ccccc}
    \toprule
  \textbf{Framework}  &  \textbf{Top-1} & \textbf{Top-3} & \textbf{Top-5} & \textbf{Top-7} \\ \midrule
  \textbf{PYInfer}&  63.034\%      & 77.931\%      & 83.774\%      & 88.865\%      \\
  \textbf{probPY}  &  58.155\%      & 75.930\%      & 79.090\%      & 80.310\%      \\ \bottomrule
  \end{tabular}
  \end{table}

In table~\ref{tab:xu16}, we present the testing recalls of PYInfer compared with the probPY model.
We evaluate PYInfer on Top-k testing recall, where $k \in \{1, 3, 5, 7\}$. 
PYInfer on Top-k returns $k$ annotations with the highest probability. 
Since we have the ground truth for annotation, 
the $|\mathcal{D}({\mathcal{X}_i})|$ is always a single type annotation. 
The $R_{\mathcal{X}_i}$ increases if the ground truth is included in the first $k$ inference
results $TOP_k$ ranked by probabilities. 
As only the Top-1 variable annotations are provided to the user side, it matters the most in real-world scenarios. 
We also collect the Top-1 precision using probPY with the setting HIGH 
probability threshold 0.95 and naming convention probability threshold = 0.7, as illustrated in their paper~\cite{xu2016python}.
We use the PYInfer model trained with the typePY dataset classifying 500 types,
and evaluate on probPY dataset. Since we did not retrain or fine-tune our model in the probPY dataset,
the performance is less competitive than our testing results. 
Still, our model outperforms the probPY model on all the Top-k recalls. 
This superiority is because we have a sufficiently large dataset for PYInfer to analyze contextual semantics.
Our embedding approach is more advanced in resolving OOV compared with word embeddings. 

Instead of assigning weights to multiple factors, our PYInfer framework is 
more efficient compared with the probPY model with the help of its parallelizable nature. 
PYInfer also outperforms the probPY on annotation coverage. 
The tool in probPY runs 68 minutes on an eight-core CPU
and generates 22,354 annotations. 
Our PYInfer model can take advantage of the GPU to run in parallel. 
It takes 113 seconds on deriving 115,535 annotations in the same dataset,
and it achieves 187.4X faster than probPY with 5.2X coverage.

\begin{table}[!htb]
  \centering
  \caption{Time and coverage comparison between PYInfer, PySonar2, and probPY Model.}
  \label{tab:timeAndCoverage}
  \begin{tabular}{lccc}
      \toprule
      \textbf{}                         & \textbf{PYInfer} & \textbf{PySonar2} & \textbf{probPY} \\ \midrule
      \textbf{Valid Annotation}        & 115,535 & 23,107   & 22,354  \\
      \textbf{Time Consumption (s)}  & 112.575 & 40       & 4,080   \\
      \textbf{Time per Annotation (ms)} & 0.974   & 1.731    & 182.518 \\ \bottomrule
    \end{tabular}
  \end{table}

Our model also proficient with higher coverage compared with PySonar2. 
Pysonar2 provides 102,361 initial annotations for the probPY dataset. 
However, many of the type annotations are question marks or invalid. 
Among 102,361 annotations, there are 38,137 question mark annotations, 
9,454 annotated as ``None'', and 40,400 without a letter (e.g., ``[[?]]''). 
Among all the question marks annotations, 
35,882 annotations can be predicted by PYInfer, 
PYInfer can also infer 9,352 valid annotations for ``None Type'' and 38,019 
annotations for ``No Letter''. 
We finally obtain 23,107 annotations after deduplicating and eliminating None and invalid types. 
Overall, our model accomplishes the best annotation coverage among all these three tools.
We reveal detailed comparisons in Table~\ref{tab:timeAndCoverage}.


\subsubsection{Comparison with TypeWriter}

TypeWriter~\cite{pradel2019typewriter} effectively leverages neural networks to infer function-level types, 
i.e., argument types and return types, from partially annotated code bases. 
It utilizes LSTM on type hints from source code tokens in argument names, usages,
and function-level comments. 
For the dataset, TypeWriter uses an internal code base as well as mypy dependencies on GitHub.
It processes 1,137 GitHub repositories and predicts 16,492 annotations for return types 
and 21,215 annotations for arguments. 

Compared with TypeWriter, we target slightly different problems and apply different frameworks.
TypeWriter infers function arguments and return types, while PYInfer annotates Python variables.
At the approach level, TypeWriter adopts traditional token-wise Word2Vec embedding,
while PYInfer applies the BPE embeddings to capture contextual code semantics. 
For neural network design, TypeWriter utilizes LSTM models on source code tokens and function comments,
while PYInfer exploits the GRU network with the attention mechanism to address local semantics. 

We compare our PYInfer model with TypeWriter in TypeWriter's testing set on generating annotations 
for three different problems: variables, function arguments, and return types,
as shown in Table~\ref{tab:TypeWriterCompare}.

\begin{table}[!htb]
\caption{Comparison between TypeWriter and PYInfer.}
\label{tab:TypeWriterCompare}
\centering
  \begin{tabular}{llrr}
  \toprule
  Top-1              &           & \textbf{PYInfer} & \textbf{TypeWriter} \\ \midrule
                    & Precision & 79\%    & 59\%       \\
                    \textbf{Variable Annotations} & Recall    & 81\%    & 47\%       \\
                    & F-1 Score & 80\%    & 52\%       \\ \midrule
                    & Precision & 72\%    & 58\%       \\
                    \textbf{Function Argument}  & Recall    & 75\%    & 50\%       \\
                    & F-1 Score & 73\%    & 54\%       \\ \midrule
                    & Precision & 59\%    & 69\%       \\
                    \textbf{Function Return}    & Recall    & 60\%     & 61\%       \\
                    & F-1 Score & 59\%     & 65\%       \\ \bottomrule
  \end{tabular}
\end{table}

\begin{itemize}[leftmargin=*]

\item \textbf{Variable Annotations.} We evaluate TypeWriter on our variable annotation dataset,
typePY. We put the variable annotations as function argument annotations in TypeWriter's model.
We re-train the TypeWriter and obtain 59\% on precision, which is similar to the performance 
of inferring argument types in TypeWriter paper. 

\item \textbf{Function Argument Types.} We evaluate PYInfer on the open-source dataset in TypeWriter.
  We put the argument name as variable names in our model. Without re-train or fine-tuning, 
  our model achieves 72\% in precision. 

\item \textbf{Function Return Types.} We evaluate PYInfer on the open-source dataset in TypeWriter.
  We treat the line of code where the function return is defined as the variable name in our model. 
  We achieve 59\% precision without re-train or fine-tuning the PYInfer model. 

\end{itemize}

From the evaluation above, our work is complementary to TypeWriter. 
TypeWriter employs the global function-level features, i.e., function source code, comments, 
and argument usages to infer return and argument types. 
Those global features provide a comprehensive view over a whole function, 
making it more advanced for inferring \emph{function-level types}. 
TypeWriter shows enhanced performance compared with NL2Type~\cite{malik2019nl2type}
and DeepTyper~\cite{hellendoorn2018deep} on argument prediction.
However, it is difficult for TypeWriter to annotate variables within a function body,
 where variable-level information is more effective. 
PYInfer annotates variables using source code semantics within a certain margin, 
making it more competitive in providing \emph{variable-level annotations}. 
The main reason is that the PYInfer model exploits the local variable-level features,
i.e., variable names and the contextual semantics within a certain margin. 
For inferring types of variables and arguments, local features are more critical since
they characterize how a variable is defined and used.

\subsection{RQ 2: Basic Types or More Types} \label{types}

We investigate the model performance of classifying only basic types in Python.
Specifically, we inspect the following built-in types: [str, int, dict, bool, float, list, tuple,
object, complex, set, type]. We analyze the distribution of different types, 
as shown in Table~\ref{tab:baseTypeDist}. 

  \begin{table}[!htb]
    \centering
    \caption{The annotation distribution on basic types with contextual data in typePY dataset.}
    \label{tab:baseTypeDist}
    \setlength{\tabcolsep}{1mm}{
    \begin{tabular}{ccc|ccc}
      \toprule
    \textbf{Type} & \textbf{Count} & \textbf{Deduplication} &  \textbf{Type} & \textbf{Count} & \textbf{Deduplication} \\ \midrule
    str    & 6,553,418         & 921,471                      & tuple                             & 78,947            & 11,110             \\
    int    & 4,028,405         & 628,552                      & object                            & 27,251            & 6,110                   \\
    dict   & 1,810,145         & 336,374                      &complex                           & 11,177            & 2,295               \\
    bool   & 4,429,648         & 241,804                      & set    & 483               & 11                   \\
    float                             & 765,077           & 126,005                      &  type   & 168               & 3                   \\
    list   & 363,832           & 48,400                       \\ \bottomrule
    \end{tabular}}
    
    \end{table}

To analyze the performance of PYInfer on basic types, we obtained 2,322,135 annotations after deduplication. 
The comparison results between basic types and 500 most common types can be found in Table~\ref{tab:basictype}.
The performance of the basic type model exceeds the 500 types model by 9.9\% on testing accuracy.  
Classifying 11 types would be much more manageable and is more precise than 500 types. 
However, a large number of user-defined types cannot be predicted with 11 types.
A model considering 500 types covers more user-defined types, making it generalizable for real-world scenarios.

  \begin{table}[!htb]
    \centering
    \caption{Comparison between basic types and the 500 most common types using PYInfer.}
    \label{tab:basictype}
    \setlength{\tabcolsep}{0.6mm}{
    \begin{tabular}{ccccccc}
      \toprule
    \textbf{Types}         & \textbf{Training} & \textbf{Validation} & \textbf{Testing} & \textbf{Precision} & \textbf{Recall} & \textbf{F-1 Score} \\ \midrule
    \textbf{Basic Types}   & 91.440\%       & 91.280\%            & 91.187\%      & 95.366\% &  91.638\%  &  93.465\%  \\
    \textbf{500 Types} & 81.165\%       & 81.153\%            &  81.195\% & 79.318\% & 81.195\% & 80.246\%     \\ \bottomrule
    \end{tabular}}
    \end{table}

\subsection{RQ 3: Threshold}


As the model provides the type annotation with a probability, 
we test different thresholds on the confidence level.
From 0.1 to 0.9, we provide results on the number of annotations with precision,
recall, and f-1 score in Table~\ref{tab:threshold} evaluated on our validation corpus in typePY. 
The annotations in Table~\ref{tab:threshold} indicates the number of variables that
are predicted with one type. With the increases in the threshold, 
our model gives out fewer annotations, while the precision increases. 
With a higher threshold, fewer annotations are given out. 
The threshold on 0.9 provides 97.677\% precision, which also neglects 42.151\% of our validation data. 

The threshold is a true reflection on the model's results on probabilities, 
which indicates how confident our model is to provide a type annotation. 
We can set a threshold to a reasonable value to attain a trade-off between 
the number of annotations we would like to obtain, and an absolute accuracy standard we would like to have.

\begin{table}[!htb]
  \center
  \caption{Evaluation results on PYInfer using typePY validation dataset with different threshold settings. }
  \label{tab:threshold}
    \begin{tabular}{cccccc}
      \toprule
    \textbf{Threshold} & \textbf{Annotations} & \textbf{Precision} & \textbf{Recall} & \textbf{F-1 Score} \\ \midrule
    \textbf{0.0}         & 699,913              & 79.510\%           & 81.275\%        & 80.383\%           \\
    \textbf{0.1}       & 689,916              & 80.339\%           & 81.185\%        & 80.760\%           \\
    \textbf{0.2}       & 677,457              & 81.556\%           & 80.880\%        & 81.217\%           \\
    \textbf{0.3}       & 653,684              & 83.596\%           & 79.944\%        & 81.729\%           \\
    \textbf{0.4}       & 617,740              & 86.631\%           & 78.029\%        & 82.105\%           \\
    \textbf{0.5}       & 578,565              & 89.663\%           & 75.340\%        & 81.880\%           \\
    \textbf{0.6}       & 540,766              & 92.146\%           & 72.208\%        & 80.968\%           \\
    \textbf{0.7}       & 502,896              & 94.118\%           & 68.508\%        & 79.296\%           \\
    \textbf{0.8}       & 460,553              & 95.985\%           & 63.809\%        & 76.657\%           \\
    \textbf{0.9}       & 404,896              & 97.677\%           & 56.881\%        & 71.895\%           \\ \bottomrule
    \end{tabular}
  \end{table}


\section{Analysis} \label{Analysis}

In this section, we focus on answering the following two research questions:

\begin{itemize}
  \item RQ 4: How does contextual information affect PYInfer's performance?
  \item RQ 5: Do BPE embeddings outperform the other learning-based code embeddings?
\end{itemize}

\subsection{RQ 4: Ablation Analysis of Contextual Semantics} \label{contextual}
We are also interested in the effectiveness of contextual semantics. 
Therefore, we conduct ablation analysis on contextual data, and analyze 
PYInfer without contextual semantics. We collect the results in Table~\ref{tab:context}. 
Without the contextual factors, our model can only take advantage of variable names, 
leaving out variable usages and the logical relations within a certain margin in a context. 
The contextual information plays a pivotal role in characterizing the source code semantic. 
The scarcity of contextual semantics devastates the effectiveness of our model. 
We observe a 41\% increment in testing accuracy with the support of contextual semantics.

  \begin{table}[!htb]
    \centering
    \caption{The evaluation results for contextual data ablation experiment of PYInfer.}
    \label{tab:context}
    \setlength{\tabcolsep}{0.6mm}{
    \begin{tabular}{ccccccc}
      \toprule
    \textbf{Model}         & \textbf{Training} & \textbf{Validation} & \textbf{Testing} & \textbf{Precision} & \textbf{Recall} & \textbf{F-1 Score}\\ \midrule

    \textbf{Context} & 81.165\%       & 81.153\%            &  81.195\% & 79.318\% & 81.195\% & 80.246\%  \\
    \textbf{No Context}   & 40.823\%       & 39.918\%       & 40.195\%     & 33.106\% &  40.775\%  &  36.542\% \\ \bottomrule

    \end{tabular}}
    \end{table}

The performance deterioration can also be partially interpreted due to the reduction of annotations.
For the model with contextual data, 
we have 3,499,933 annotations considering 500 types.
Without contextual information,
we have an overwhelming number of duplicates as most of the variables have a similar 
name with the same type. 
We only obtain 841,521 annotations after deduplication. 
As we omit the contextual data and deduplicate annotations,
we encounter a dramatic shortage of type annotations.
This insufficiency on annotations also deteriorates the model's performance.

\begin{figure}[!htb]
  \includegraphics[width=\linewidth]{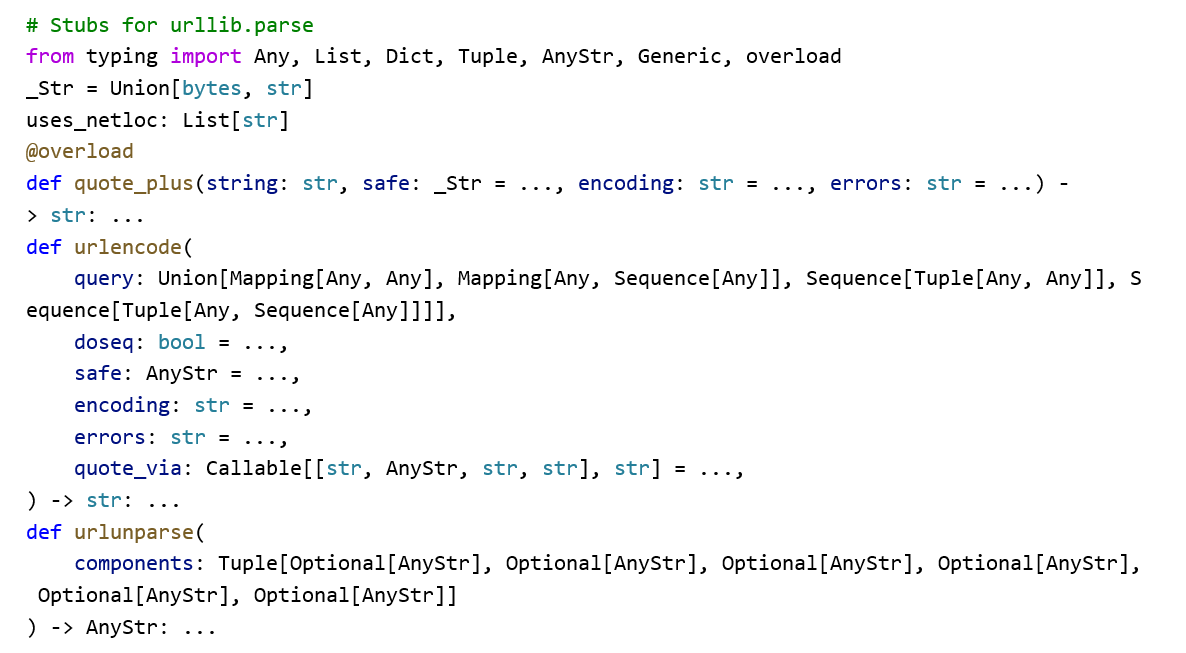}
  \caption{A pyi file sample in typeshed dataset.}
  \label{fig:pyi}
\end{figure}

To further investigate the significance of the contextual semantics, 
we conduct experiments with the human-labeled typeshed dataset. 
It contains type annotations in the format of a pyi file following particular formats~\cite{pep0484,pepF3107}, 
which is widely used for type checking and type inference. 

In Fig.~\ref{fig:pyi}, 
we present a sample pyi file in the typeshed GitHub repository. 
A pyi file contains annotations on parameters and return values for functions.
Note that for the variable \emph{safe} in Fig.~\ref{fig:pyi}, 
we are provided with both \emph{\_Str} and \emph{AnyStr} annotations under different contexts. 
Since source code is unavailable for this dataset, 
we are unable to separate these two variables with corresponding annotations.

\begin{figure*}[!htb]
  \centering
  \includegraphics[width=\linewidth]{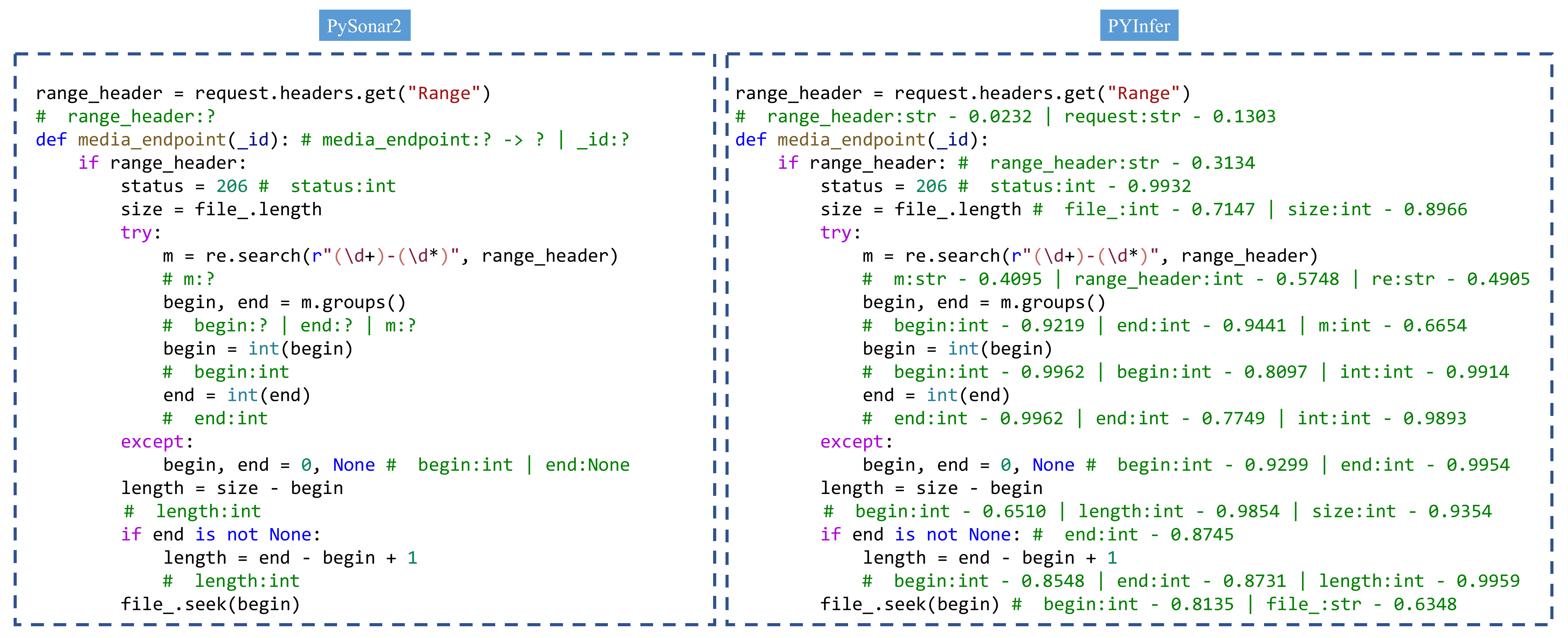}
  \caption{We compare the type annotations provide by PySonar2 (left) and PYInfer (right). 
  PYInfer not only generates more annotations for complex scenarios but also provides the probabilities for each type.}
  \label{fig:PySonar2VSPYInfer}
\end{figure*}

The number of annotations in the typeshed dataset is extremely limited.
There would be a tremendous data shortage for each type when considering 500 types. 
As a mitigation, we are devoted to investigating the top-50 types.
We retrain our PYInfer on typeshed and compare the model trained on the typePY dataset.
PYInfer trained on typeshed achieves 41.747\% testing accuracy,
 53.126\% precision, 43.748\%, and 47.983\% f-1 score.
Compared with the results in Table~\ref{tab:context}, 
we find a similar trend on the testing side, 
which points out that the contextual information is critical for variable type prediction. 

\subsection{RQ 5: Advantages of Source Code Embeddings} \label{glove}
Besides the contextual data and types, our model is also superior 
in adopting the BPE embeddings. 
Source code is rich in user-defined variables and function names. 
We often encounter the OOV issue in characterizing source code semantics. 
We apply the BPE method to address this problem,
which enables us to make full use of the contextual code semantics.

In this RQ, we mainly investigate the effectiveness of the BPE embeddings 
compared with learning-based GloVe embeddings, a prominent word embedding approach for text files. 
We retrain our GloVe embeddings on our typePY dataset and derive a 50-dimensional vector for each token. 
Each contextual token within a certain margin is mapped to a 50-dimensional GloVe vector. 
We then concatenate the embedding vectors for tokens before and after the current 
variable within a certain margin, and the name of the current variable. 
The concatenated vector is fed through the GRU network for model training. 
We collect and gather the results in Table~\ref{tab:GloVe}. 
Our experiment results reveal that the BPE embeddings outperform the GloVe embeddings significantly.
We demonstrate that BPE is more effective than GloVe embeddings for variable typing.

  \begin{table}[!htb]
    \centering
    \caption{Comparison between BPE embeddings and GloVe embeddings with PYInfer on typePY dataset.}
    \label{tab:GloVe}
    \setlength{\tabcolsep}{0.5mm}{
    \begin{tabular}{ccccccc}
      \toprule
    \textbf{Embeddings} & \textbf{Training} & \textbf{Validation} & \textbf{Testing} &  \textbf{Precision} & \textbf{Recall} & \textbf{F-1 Score} \\ \midrule
    \textbf{BPE}       & 81.165\%       & 81.153\%    &  81.195\%  & 79.318\% & 81.195\% & 80.246\% \\
    \textbf{GloVe}     &  51.916\%       & 54.168\%   & 54.077\%   & 50.470\% & 54.132\% &  52.237\%   \\ \midrule
    \end{tabular}}
    \end{table}

\section{Discussion} \label{Discussion}

\subsection{Strengths and Weaknesses of PYInfer}

The main reason why PYInfer outperforms the other type inference
tools is that we have a sufficient large annotated dataset,
 and encode the contextual semantics of a variable into
the deep learning model.
As illustrated in Fig.~\ref{fig:margin}, we set up a margin to define the range
of contexts, adopt BPE embeddings to resolve OOV, 
and employ GRU network with attention mechanism on semantic extraction. 
The margin settings influence PYInfer's performance.
We conduct experiments to examine the influence on margin setting based on the probPY dataset,
and the evaluation results can be found at Table~\ref{tab:xu16margin}.
The experiments are performed on merged probPY dataset evaluating Top-k
 ($k\in \{1, 3, 5, 7\}$) without threshold.
As the margin increases, the accuracy increases first and then decreases, 
which can be explained in the following ways: 
The margin settings on 32 and 64 lack adequate contextual local semantics for the current variable. 
The margin settings of 256 and 512 provide more irrelevant contextual semantics. 
We derive the following insight that a proper setting of the margin is undoubtedly 
influential to our model's overall performance.

\begin{table}[!htb]
  \centering
  \caption{Different margin settings with PYInfer on probPY dataset.}
  \label{tab:xu16margin}
  \begin{tabular}{cccccc}
    \toprule
  \textbf{Margin}    & 32       & 64       & 128      & 256      & 512      \\ \midrule
  \textbf{Accuracy} & 54.751\% & 60.055\% & 61.635\% & 60.933\% & 59.265\%  \\
  \textbf{Precision} & 69.446\% & 69.016\% & 68.554\% & 67.862\% & 65.972\% \\
  \textbf{Recall}  & 54.751\%  &  60.055\% & 61.635\% & 60.933\% & 59.265\% \\
  \textbf{F-1 Score} & 61.229\% & 64.224\% & 64.911\% & 64.211\% & 62.439\% \\ \bottomrule

  \end{tabular}
  \end{table}


  There are several limitations of PYInfer. 
  Although PYInfer is qualified to process user-defined types, 
  several annotations are incorrect due to the limited training data on 
  user-defined types compared with built-in types. 
  We have shown the distribution of different annotations in Table~\ref{tab:P1DataDist},
  and the number of samples on user-defined types are overwhelmed by the number of built-in types.
  Also, as many static analyzers, PYInfer demands access to source code, 
  which might be unrealistic due to confidential issues.

  \subsection{Superiority over PySonar2}

  We have to emphasize that although it is trained using the annotations generated by PySonar2, 
  our model shows remarkably higher coverage than Pysonar2 on variable type inference. 
  Analyzing the same code snippet\footnote{Accessed 
  from \url{https://github.com/pyeve/eve/blob/master/eve/endpoints.py}.}, 
  we compare the annotation results of PySonar2 and 
  PYInfer trained with the 500 most common types on the typePY dataset, as shown in Fig.~\ref{fig:PySonar2VSPYInfer}. 
  PySonar2 and PYInfer have both reported several false positives. 
  PySonar2 utilizes control-flow aware interprocedural analysis, 
  which leaves out the semantic knowledge in source code. 
  It annotates half of the variables with question marks 
  (7 out of 14 annotations). 
  PYInfer captures the code semantics by considering contextual semantics 
  within a specific margin, which gives out specific annotations when the contexts are different. 
  It can deal with demanding scenarios where PySonar2 emits question marks, and
   correctly annotates 21 variables within all 29 annotations. 
  Although PYInfer is not sound, it achieves reasonable testing accuracy 
  (91.187\% on 11 basic types, and 81.195\% on 500 most common types).



\section{Related Work}


\subsection{Type Inference for Python}

Standards, such as PEP 484~\cite{pep0484}, PEP 3107~\cite{pepF3107}, 
are proposed to facilitate type hints and annotations. 
Type checkers, such as mypy~\cite{mypy}, pyre-check~\cite{pyre-check}, pytype~\cite{pytype},
and pyright~\cite{pyright}, take advantage of annotations to detect type inconsistencies. 
Without the type annotations, it would be futile for type checkers to detect 
type inconsistencies for Python projects. 
Existing type checkers primarily rely on manually written type annotations from developers, 
which are expensive to provide. 

Previous type inference work often leaves out the natural language elements in source code. 
Several existing type annotators adopt dynamic analysis. PyAnnotate~\cite{pyannotate}, 
for example, is a dynamic type inference tool developed by Dropbox. 
Using dynamic analysis to acquire type annotation during runtime makes the inference 
accurate for specific input. However, it requires a particular runtime environment, 
which is not realistic under some real-world circumstances. Also, it is input sensitive, 
which leads to limited coverage. 
The frameworks proposed by~\citet{cannon2005localized, salib2004starkiller, vitousek2014design} 
generate type annotations based on static analysis. 
\citet{salib2004starkiller} presented a static type inference and compiler 
for Python with a modified Cartesian Product Algorithm~\cite{agesen1995cartesian}. 
This tool converts Python source codes into equivalent C++ codes, 
making it proficient in analyzing codes in a foreign language. 
Nevertheless, this tool is not complete, and the instructions that can be processed are limited.
\citet{vitousek2014design} developed gradual typing for Python using a type system
based on first-order object calculus~\cite{abadi1996theory} augmented with dynamic types. 
\citet{luo2018recognizing} analyzed Python docstrings and
built a decision tree classifier. 
This approach considers only nine built-in types, and will be futile if docstring is unavailable.
PySonar2\cite{pysonar2}, a static type inference tool for Python, 
can statically generate type hints for variables.
Although it only covers 48.91\% annotations based on our experiments,
the accurate results accommodate our research with ground-truth for the labeling.

Several related works exploit semantics in source code using neural networks. 
\citet{pradel2019typewriter} predict the argument and return types for functions 
with natural language factors in source code and comments. 
It is more effective on function-level inference compared with variable-level prediction. 
\citet{xu2016python} adopt the probabilistic inference to derive type annotations for Python. 
Considering various probabilistic factors makes the tool time-consuming, and it is hard to be scalable.

\subsection{Type Inference on Dynamically Typed Languages}

Existing learning-based type inference tools mainly focus on JavaScript, 
where the type annotations can be obtained using TypeScript~\cite{bierman2014understanding}. 
LambdaNet~\cite{wei2020lambdanet} utilizes graph neural networks to predict types with 
contextual hints involving naming and variable usage. 
It defines the type dependency graph and 
propagates type information on a graph neural network. 
LambdaNet explores the potential of using graph embeddings on type inference.
Compared with graph-based embeddings, generating token-based source code embedding is more efficient.
It can exploit source code semantics and more accessible to be applied to other languages. 
NL2Type~\cite{malik2019nl2type} proposes a learning-based approach to predict 
type signatures for functions with natural language support. 
\citet{hellendoorn2018deep} use a deep learning model with 300-dimensional word embeddings
to generate type annotations for JavaScript.

\subsection{Learning-based Source Code Analysis}

Machine learning has been widely adopted in program analysis. 
\citet{godefroid2017learn} proposed a statistical machine learning
technique to generate grammar-suitable input for fuzzing automatically. 
DeepFix~\cite{gupta2017deepfix} uses a sequence-to-sequence neural network 
with attention to detect and fix errors in C programs. 
\citet{raychev2014code} focused on code completion with APIs using a language model. 
Deepsim~\cite{zhao2018deepsim} measures code similarity using a deep learning model 
with the control flow and data flow matrix. 
\citet{raychev2015predicting} built JSNice to predict the names of identifiers and 
type annotations of variables for JavaScript. 
\citet{pradel2018deepbugs} proposed a learning-based approach to detect 
three different kinds of name-based bugs. 
They generate negative samples using code transform based on specific bug patterns. 
Extracting bug patterns and seeding bugs into source code following the pattern demands great human efforts.

\subsection{Learning-based Source Code Embeddings}

Learning-based word embeddings have been widely adopted in natural language
processing, such as word2vec~\cite{mikolov2013distributed}, and
doc2vec~\cite{le2014distributed}. \citet{nguyen2017exploring} took advantage
of word2vec to analyze semantic relations on API usages. \citet{ye2016word} 
trained word embeddings on API documents, tutorials, 
and reference documents to estimate semantic similarities. 
These approaches highly rely on the training data, 
and the tokens not observed in training inputs are not well addressed. 
OOV is a severe problem, especially in source code embeddings, 
as a programmer can flexibly define function and variable names. 
BPE algorithm~\cite{sennrich2015neural} addresses the OOV problem by encoding rare 
and unknown words as sequences of subword units. 
It is effective in 
many program analysis tasks~\cite{babii2019modeling,karampatsis2019maybe,karampatsis2020big}. 

Besides token-based source code embeddings, 
there are also graph-based embeddings. 
Code2Vec~\cite{alon2019code2vec}  decomposes source code into a 
collection of paths in its abstract syntax tree (AST) and learns 
the atomic representation of each path. 
\citet{alon2018general} proposed a path-based representation 
for learning from programs using the AST. 
LambdaNet~\cite{wei2020lambdanet} proposes the type dependency graph, 
which links type variables with logical constraints as well as name and usage information. 
These embeddings primarily consider structures in source code. 
We investigate the source code semantics in variable names and contextual information, 
which is different from graph-based code structure with AST and graph neural networks. 

\section{Conclusions}
We have presented PYInfer, a learning-based approach to generate type
annotations for Python automatically. 
Our main contribution is the end-to-end PYInfer framework to infer variable types for Python statically.
Our research combines symbolic and probabilistic methods to generate 
type annotations for variables, which turns out to be highly effective and efficient. 
One of our key insights is to consider
contextual information for variables, which enhances the model
by encoding additional code semantics.  PYInfer is capable of
handling user-defined types by formulating the type inference as a
classification problem. It achieves 5.2X more on code coverage and
187X faster than a state-of-the-art technique for Python type inference, and
covers 5X more variables compared to PySonar2. 
It outperforms TypeWriter on inferring types for variables and function arguments.
Finally, we propose a method of data collection and contribute a large dataset
consisting of 77,089,946 type annotations from 4,577 popular Python projects. We
make our tool and datasets publicly available to facilitate further research.


\bibliography{IEEEabrv,ref}

\end{document}